\documentclass[aps,prl,groupedaddress]{revtex4}
\usepackage{graphicx}

\def\be{\begin{equation}}
\def\ee{\end{equation}}
\def\ba{\begin{eqnarray}}
\def\ea{\end{eqnarray}}
\def\bc{\begin{center}}
\def\ec{\end{center}}
\def\sgn{{\rm sgn}}
 
\begin{document}

\title{Non-linear electromagnetic response of graphene}

\author{S. A. Mikhailov}

\affiliation{Institute for Theoretical Physics II, University of Augsburg, D-86135 Augsburg, Germany}

\date{\today}

\begin{abstract}
It is shown that the massless energy spectrum of electrons and holes in graphene leads to the strongly non-linear electromagnetic response of this system. We predict that the graphene layer, irradiated by electromagnetic waves, emits radiation at higher frequency harmonics and can work as a frequency multiplier. The operating frequency of the graphene frequency multiplier can lie in a broad range from microwaves to the infrared.
\end{abstract}


\maketitle

In the past two years a great deal of attention has been attracted by a recently discovered, new two-dimensional (2D) electronic system -- graphene, built out of a single monolayer of carbon atoms with a honeycomb 2D crystal structure \cite{Novoselov05,Zhang05}. The band structure of the charge carriers in this system consists of six Dirac cones at the corners of the hexagon-shaped Brillouin zone \cite{Wallace47,Semenoff84}, with the massless, linear electron/hole dispersion. The massless electron spectrum leads to unusual transport and electrodynamic properties, which have been intensively studied in the literature, see e.g. 
\cite{Novoselov07,Gusynin05,Ziegler06,Katsnelson06,Cheianov06,Nilsson06,Tudorovskiy06,Nomura07,Fuchs07,Bostwick07,Deacon07,Gusynin06a,Falkovsky06,Sadowski06,Sadowski07,Gusynin06b,Gusynin07,Gusynin07a,Hwang06,Vafek06,Apalkov06,Abergel06,Ryzhii06,Trauzettel07,Ryzhii07,Rana07,Mikhailov07} 
 and for review \cite{Katsnelson07,Geim07a}. 

Electrodynamic properties of graphene have been theoretically studied in Refs. 
\cite{Gusynin06a,Gusynin06b,Gusynin07,Gusynin07a,Falkovsky06,Sadowski06,Sadowski07,Hwang06,Vafek06,Apalkov06,Abergel06,Ryzhii06,Trauzettel07,Ryzhii07,Rana07,Mikhailov07}. The frequency dependent conductivity\cite{Gusynin06a,Gusynin06b,Gusynin07,Gusynin07a,Falkovsky06}, as well as plasmon \cite{Hwang06,Apalkov06,Ryzhii06,Ryzhii07,Rana07}, plasmon-polariton \cite{Vafek06}, and transverse electromagnetic wave spectra \cite{Mikhailov07} have been investigated. In all these papers electrodynamic response of the system has been studied within the {\em linear response} theory (for instance, using the Kubo formalism, or the random phase approximation, or the self-consistent-field approach). In this Letter we show that, apart from all the fascinating and non-trivial properties of graphene predicted and observed so far, this material should also demonstrate  {\em strongly non-linear} electrodynamic behavior. In particular, irradiation of the graphene sheet by a harmonic electromagnetic wave with the frequency $\Omega$ should lead to the emission of the higher harmonics with the frequencies $m\Omega$, $m=3,5,\dots$, from the system. The operating frequency of such a frequency multiplier can vary from microwaves up to infrared, and the required ac electric field is rather low, especially at low carrier densities and low temperatures. The predicted non-linear electrodynamic properties of graphene may open up new exciting opportunities for building electronic and optoelectronic devices based on this material.

To qualitatively demonstrate the non-linear behavior of graphene electrons consider a classical 2D particle with the charge $-e$ and the energy spectrum $\epsilon_{\bf p}=V p=V\sqrt{p_x^2+p_y^2}$ in the external electric field $E_x(t)=E_0\cos\Omega t$. Here $V$ is the velocity of 2D electrons in the energy band (in graphene $V\approx 10^8$ cm/s \cite{Novoselov05,Zhang05}). According to the classical equations of motion $dp_x/dt=-eE_x(t)$ the momentum $p_x$ will then be equal to $p_x(t)=-(eE_0/\Omega)\sin\Omega t$, and the velocity $v_x=\partial \epsilon_{\bf p}/\partial p_x$ is then $v_x(t)=-V\sgn(\sin\Omega t)$. If there are $n_s$ particles per unit area, the corresponding ac electric current 
\be
j_x(t)= en_sV \sgn (\sin \Omega t)=en_sV \frac 4\pi\left\{\sin\Omega t + \frac 13 \sin 3\Omega t + \frac 15 \sin 5\Omega t + \dots\right\}\label{estim}
\ee
contains all odd Fourier harmonics, with the amplitudes $j_m$, $m=1,3,5\dots$, falling down very slowly with the harmonics number, $j_m\sim 1/m$. Notice that at the density $n_s=6\cdot 10^{12}$ cm$^{-2}$ and at $V\simeq 10^8$ cm/s (parameters of Refs. \cite{Novoselov05,Zhang05}) the current amplitude $j_0=en_sV$ in our simple estimate gives a giant value of $j_0\simeq 100$ A/cm. 

The above consideration does not take into account the Fermi distribution of charge carriers over the quantum states in the conduction and valence bands of graphene. To get a more accurate description of the non-linear phenomena in the considered system we use the kinetic Boltzmann theory, which allows one to get an exact response of the system not imposing any restrictions on the amplitude of the external ac electric field ${\bf E}(t)$. Using this quasi-classical approach we take into account the intra-band contribution to the ac electric current. The inter-band contribution to the electric current, due to the transitions between the hole and the electron bands, is ignored. This imposes certain restrictions on the frequency of radiation $\Omega$, which will be discussed below.

Consider a 2D electron/hole gas with the energy spectrum $\epsilon_{{\bf p}\pm}=\pm V\sqrt{p_x^2+p_y^2}$ under the action of the field ${\bf E}=(E_x,0)$, where the sign $+$ ($-$) corresponds to the electron (hole) band, $E_x(t)=E_0e^{\alpha t} \cos(\Omega t)$, and $\alpha\to +0$ describes an adiabatic switching on of the electric field. Assume that the Fermi energy $\epsilon_F$ lies in the electron (or the hole) band and that the temperature is small as compared to $\epsilon_F$, $T\ll \epsilon_F$. The momentum distribution function of electrons $f_{{\bf p}+}(t)\equiv f_{\bf p}(t)$ (we omit the sign $+$ for brevity) in the collisionless approximation is described by the Boltzmann equation
\be
\frac{\partial f_{\bf p}(t) }{\partial t} -\frac{\partial f_{\bf p}(t)}{\partial p_x} e E_0 e^{\alpha t} \cos(\Omega t)=0,
\ee
which has the exact solution
\be
f_{\bf p}(t) = F_0\left(p_x-p_0(t),p_y\right),
\ee
where 
\be
F_0(p_x,p_y) =\left[1+\exp\left(\frac{ V\sqrt{p_x^2+p_y^2}-\epsilon_F}T\right)\right]^{-1}
\ee
is the electronic Fermi function, and $p_0(t)=-(eE_0/\Omega)e^{\alpha t}\sin\Omega t$ is the solution of the single particle equation of motion. The electric current ${\bf j}(t)=-eg_sg_vS^{-1}\sum_{{\bf p}} {\bf v} f_{\bf p}(t)$ then assumes the form
\be
j_x(t)= -\frac{g_sg_veV}{(2\pi \hbar)^2}\int \int dp_xdp_y \frac{p_x}{\sqrt{p_x^2+p_y^2}} F_0\left(p_x-p_0(t),p_y\right),
\label{jx1}
\ee
$j_y=0$, where $g_s=g_v=2$ are the spin and valley degeneracies in graphene, and $S$ is the sample area. After some lengthy but simple transformation, Eq. (\ref{jx1}) can be rewritten as 

\be
j_x(t) = en_sV \frac 4\pi \frac{2Q_F(t)}{\sqrt{1+Q_F^2 (t)}}
\int_{0}^{\pi/2} 
\frac{\cos^2 x dx}{
\sqrt{1+\frac{2Q_F(t)}{1+Q_F^2 (t)}\cos x} +
\sqrt{1-\frac{2Q_F(t)}{1+Q_F^2 (t)}\cos x}
}\label{jx2}
\ee
where
\be
n_s\equiv n_e=\frac{g_sg_vp_F^2}{4\pi \hbar^2}=\frac{g_sg_v\epsilon_F^2}{4\pi \hbar^2 V^2}
\ee
is the density of electrons, $p_F=\epsilon_F/V$ is the Fermi momentum, and
\be
Q_F(t)=-\frac{p_0(t)}{p_F}=\frac{eE_0V}{\Omega \epsilon_F}\sin(\Omega t)\equiv Q_{F0}\sin(\Omega t)
\ee
is the field parameter, proportional to the ac electric field $E_0$.

\begin{figure}
\includegraphics[width=8.5cm]{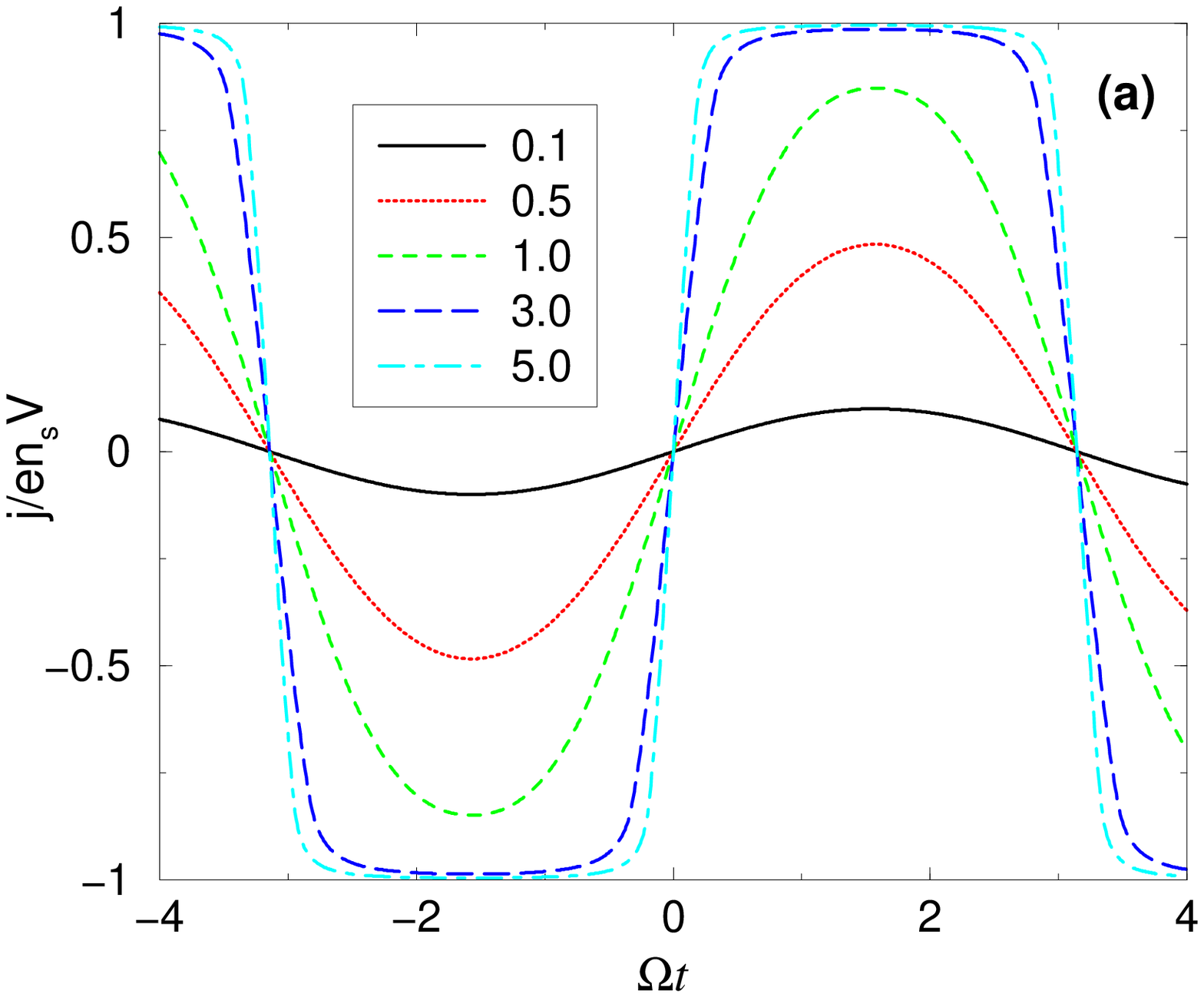}\includegraphics[width=8.5cm]{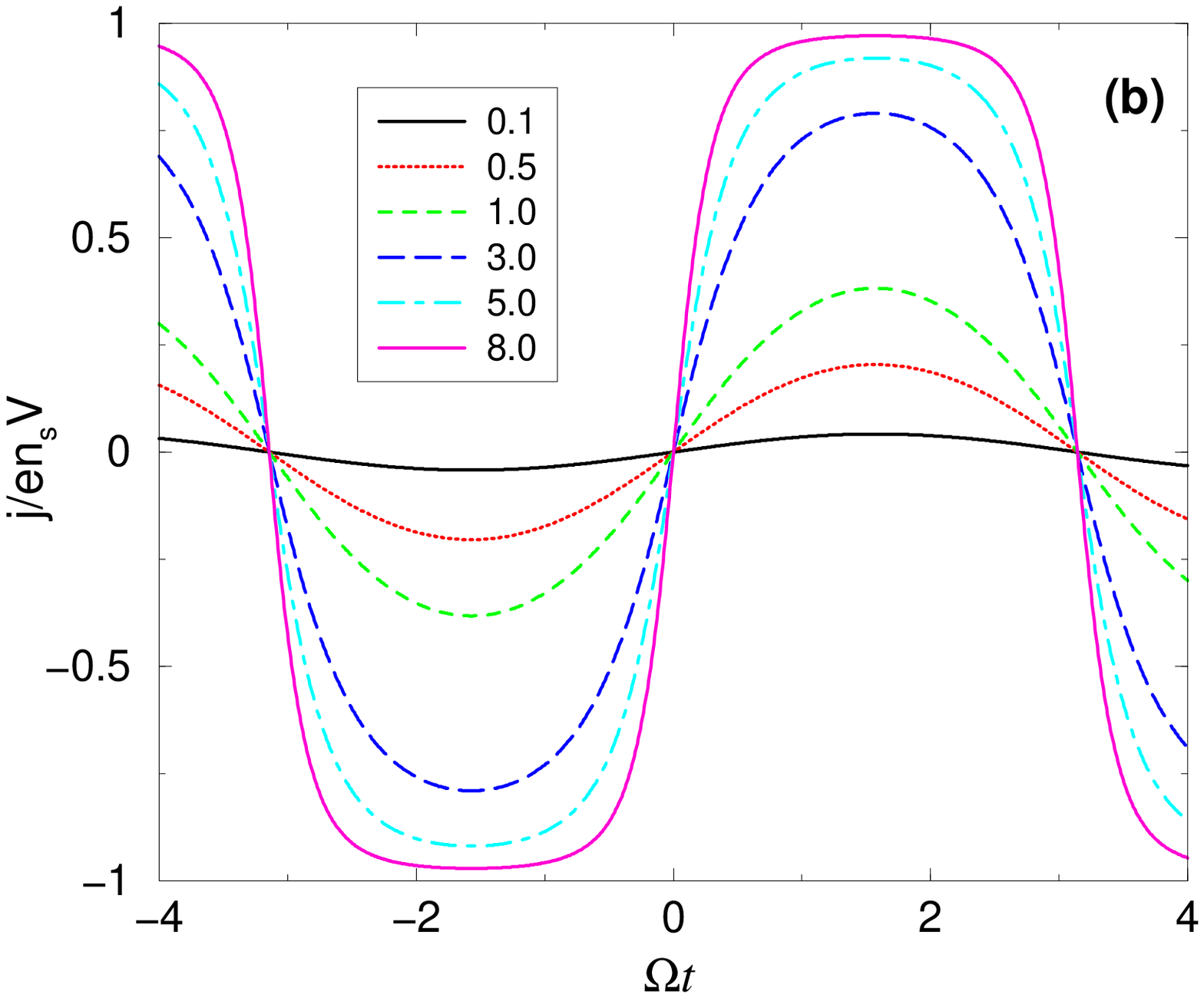}
\caption{\label{T0} (Color online) The time dependence of the ac electric current, measured in units $en_sV$, at harmonic excitation of the system at the frequency $\Omega$.  (a) The temperature is zero, $T/\epsilon_F=0$; the curves are labeled by the values of the electric field parameter $Q_{F0}=eE_0V/\Omega \epsilon_F$. (b) The temperature is finite, the Fermi energy is zero, $\epsilon_F=0$; the curves are labeled by the values of the parameter $Q_{T0}=eE_0V/\Omega T$.
}
\end{figure}

Figure \ref{T0}a shows the current (\ref{jx2}) as a function of time $\Omega t$. One sees that in the low-field limit the response is linear. Expanding the current (\ref{jx2}), we get at $Q_{F0}\ll 1$
\be
j_x(t) \approx 
en_sV Q_{F0}\left\{\left(1-\frac 3{32}Q_{F0}^2\right)\sin\Omega t+\frac 1{32} Q_{F0}^2\sin 3\Omega t\right\},
\ee
so that the linear response conductivity (in the collisionless approximation) is
\be
\sigma_{\epsilon_F,T=0}(\Omega)=\frac {in_se^2V}{\Omega p_F}=i\frac{e^2}\hbar\frac{g_sg_v}{4\pi}\frac{\epsilon_F}{\hbar\Omega}.\label{lincond}
\ee
The expression (\ref{lincond}) coincides with the intra-band Drude conductivity, which can be obtained from the linear-response theory \cite{Gusynin06a,Gusynin06b,Gusynin07,Gusynin07a,Falkovsky06,Mikhailov07}. As the inter-band conductivity is of order of $e^2/\hbar$ \cite{Gusynin06a,Gusynin06b,Gusynin07,Gusynin07a,Falkovsky06,Mikhailov07}, our quasi-classical approach is valid at $\hbar\Omega\lesssim \epsilon_F$. At the electron density $\simeq 10^{11}-10^{12}$ cm$^{-2}$ this restricts the frequency by the value of $10-30$ THz. 

In the strong-field limit  $Q_{F0}\gtrsim 1$ Eq. (\ref{jx2}) results in the formula (\ref{estim}). From the condition $Q_{F0}\gtrsim 1$, rewritten as
\be
E_0\gtrsim 
\frac{2 \hbar \Omega\sqrt{\pi n_s}}{e\sqrt{g_sg_v}},
\label{condition}
\ee
one sees that the required ac electric field grows linearly with the electromagnetic wave frequency and is proportional to the square root of the electron density. At $f\simeq 50$ GHz and $n_s\simeq 10^{11}$ cm$^{-2}$, the inequality (\ref{condition}) is fulfilled at $E_0\gtrsim 100$ V/cm. This value can be reduced in systems with lower electron/hole density. Therefore, we consider now an opposite limiting case with $\epsilon_F=0$, but finite temperature $T$.

At finite $T$ and the vanishing $\epsilon_F=0$ both electrons and holes contribute to the charge carrier density
\be
n_s=n_e+n_h=\frac{\pi g_sg_vT^2}{12\hbar^2V^2}
\ee
and to the current. Starting again from Eq. (\ref{jx1}) but accounting for the hole contribution and putting $\epsilon_F=0$, we get
\be
j_x(t)=
en_sV\frac{12}{\pi^3}\int_0^\infty xdx \int_{0}^\pi d\theta  \frac {\cos\theta}{1+\exp\left(\sqrt{x^2+Q_T^2(t)-2x Q_T(t)\cos\theta}\right)},\label{jx3}
\ee
where
\be
Q_T(t)=-\frac{V p_0(t)}{T}=\frac{eE_0V}{\Omega T}\sin\Omega t\equiv Q_{T0}\sin\Omega t.
\ee

\begin{figure}
\includegraphics[width=8.5cm]{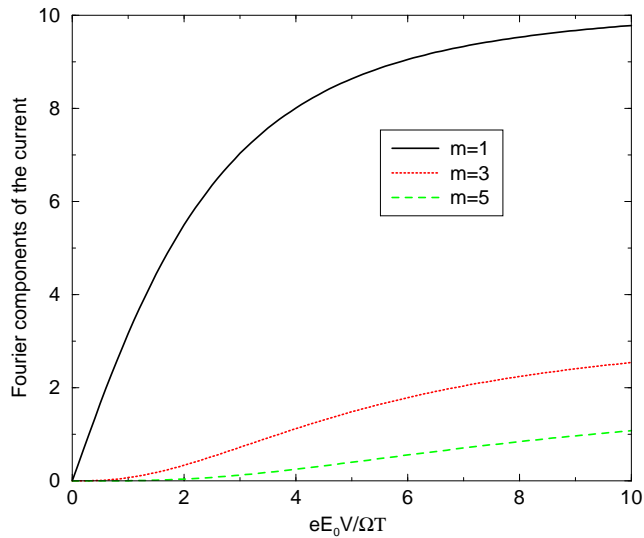}\\
\caption{\label{Jmu0Four} (Color online) The Fourier components of the current (\ref{jx3}), in arbitrary units, as a function of $Q_{T0}=eE_0 V/\Omega T$ at $\epsilon_F/T=0$. 
}
\end{figure}

Figure \ref{T0}b shows the current (\ref{jx3}) as a function of time $\Omega t$. In the low-field limit $Q_{T0}\ll 1$ we get from (\ref{jx3}) the current
\be
j_x(t)\approx en_sV Q_T(t)\frac {6\ln 2}{\pi^2},
\ee
and the correct expression for the linear-response intra-band dynamic conductivity \cite{Falkovsky06},
\be
\sigma_{\epsilon_F=0,T}(\Omega)=\frac {6\ln 2}{\pi^2}\frac{in_se^2V^2}{T\Omega}
=i\frac {\ln 2}{2\pi}\frac{e^2}\hbar\frac{g_sg_vT}{\hbar\Omega}.
\ee
One sees that the quasi-classical approach is now valid at $\hbar\Omega\lesssim T$. This restricts the frequency by the value of $\simeq 200$ GHz at $T\sim 10$ K and $\simeq 6$ THz at room temperature. 
In the strong field regime $Q_{T0}\gtrsim 1$ Eq. (\ref{jx3}) is reduced, again, to (\ref{estim}). Figure \ref{Jmu0Four} shows the Fourier components of the ac electric current, for $m=1$, 3 and 5, as a function of the field parameter $eE_0 V/\Omega T$ at $\epsilon_F=0$. The strong-field condition now assumes the form
\be
E_0\gtrsim\frac{\Omega T}{eV}.
\ee
At $T\simeq 10$ K and $f\simeq 100$ GHz this gives a moderate value of the required electric field $E_0\simeq 5$ V/cm. The efficiency of the predicted frequency multiplication effect can be increased further by using the resonance response of the system at the plasmon, the cyclotron, or the magnetoplasmon frequency.

To summarize, we have investigated the non-linear electrodynamic response of 2D electrons and holes in graphene. We have shown that irradiation of graphene by the electromagnetic wave with the frequency $\Omega$ should lead to the higher harmonics generation at frequencies $3\Omega$, $5\Omega$, et cetera. The efficiency of the frequency up-conversion is rather high: the amplitudes of the higher harmonics of the ac electric current fall down slowly (as $1/m$) with the harmonics index $m$. The presented quasi-classical theory is valid at $\hbar\Omega\lesssim \max\{\epsilon_F,T\}$. This estimate shows that the effect works at frequencies up to $5-10$ THz, which opens up exciting opportunities for building new graphene devices for terahertz and sub-terahertz electronics. 
 
I wish to thank Klaus Ziegler for stimulating discussion. The work was partly supported by the Swedish Research Council and INTAS.

\bibliography{../../BIB-FILES/mikhailov,../../BIB-FILES/lowD,../../BIB-FILES/dots,../../BIB-FILES/graphene,../../BIB-FILES/thz,../../BIB-FILES/zerores}

\end{document}